\documentclass[showpacs, twocolumn]{revtex4-1}
\usepackage{graphicx}
\usepackage{amsmath}

\begin{document}

\title{Spontaneous formation of bright solitons in self-localized impurities in Bose-Einstein condensates}

\author{Abdel\^{a}ali Boudjem\^{a}a}

\affiliation{Department of Physics, Faculty of Sciences, Hassiba Benbouali University of Chlef P.O. Box 151, 02000, Ouled Fares, Chlef, Algeria.}

\email {a.boudjemaa@univ-chlef.dz}


\begin{abstract}
We study the formation of bright solitons in the impurity component of Bose-Einstein condensate-impurity mixture by using the time-dependent Hartree-Fock-Bogoliubov theory.
While we assume the boson-boson and impurity-boson interactions to be effectively repulsive,
their character can be changed spontaneously from repulsive to attractive in the presence of strong anomalous correlations. 
In such a regime the impurity component becomes a system of effectively attractive atoms leading automatically to the generation of bright solitons.
We find that this soliton decays at higher temperatures due to the dissipation induced by  the impurity-host and host-host interactions.
We show that after a sudden increase of the impurity-boson strength a train of bright solitons is produced 
and this can be interpreted in terms of the modulational instability of the time-dependent impurity wave function. 

\end{abstract}


\maketitle

\section{Introduction} \label{intro}

A soliton is a self-focusing solitary wave that maintains its shape while it travels at constant speed and arises from a balance between nonlinear and dispersive effects. 
Bose-Einstein condensates (BECs) constitute a best environment for studying nonlinear macroscopic excitations in quantum systems.
Excitations in the form of dark solitons with repulsive interaction have been realized in \cite{Bur, Dens, Ander} one and half decade ago.

Bright solitons have been observed in BECs of  ${}^7$Li in quasi-one-dimensional (1D) regime \cite {Khay, Stre}. The observation of bright solitons was therefore possible 
only by means of the Feshbach resonance and then tuning of the interactions from repulsive to attractive during the experiments.
In the experiment of Strecker et\textit {al}.\cite {Stre}, the formation of the bright  soliton trains has been interpreted as due to
quantum mechanical phase fluctuations of the bosonic field operator \cite{usama}. 
Bright soliton trains can be also generated in a BEC embedded in a quantum degenerate Fermi gas\cite {karp} 
as a result of a competition between two interparticle interactions: boson-boson collisions which are effectively repulsive and boson-fermion collisions which are attractive.

In this paper we propose a novel scheme to realize bright solitons in quasi-1D atomic quantum gases.
In particular, we study the formation of bright solitons in the impurity component of BEC-impurity mixture at finite temperature 
by employing a versatile model known as time dependent Hartree-Fock-Bogoliubov (TDHFB)  \cite {boudj2014, boudj12014}.
Experimentally, such mixtures have been already realized with a medium composed of either bosonic \cite {Chik, Cat, Nicol, Scell}
or fermionic atoms \cite {And, Koh, Kosc}. Impurities in a Bose gas (Bose polarons) have been the subject of intense 
theoretical \cite {Tim1, Jack1, Tim2, Tim3, Tim4,Temp, Jack2, Jack3, Tim4,Tim5, Dem, Stef} studies.  
An important feature of these mixtures is that when neutral impurity atoms immersed in a BEC can spontaneously form a self-localize state.
This localized state, within the strong coupling approach, exhibits a solitonic behavior at both zero and finite temperatures \cite {Tim2, boudj2014} in quasi-1D geometry. 
These solitons are reminiscent of the well known optical wave solitons \cite {Yuri}.

Here, we have pointed out that the bright solitons can be created spontaneously in the impurity component for high anomalous density.
This latter quantifies the correlations between pairs of condensed atoms with pairs of non condensed atoms and rises with interactions strength. 
It was shown that the anomalous correlations play a crucial role on the stability of BEC and on the occurrence of the superfluidity \cite {boudj2012, boudj2012}.
We find that the single bright soliton decays at nonzero temperatures, with the decay rate increasing with rising temperature
owing to  the host-host and impurity-host interactions. 
In addition, we show that bright soliton trains can be produced automatically due to the modulational instability (MI) of the evolving classical phase 
in the impurity component of a  harmonically trapped BEC-impurity mixture even with repulsive impurity-boson and boson-boson interactions. 
These trains generate without changing the trap geometry as has been suggested in \cite {karp}, neither using the Feshbach resonance as it has been observed in \cite {Khay, Stre}, 
or even without imprinting the initial wave function with a fluctuating phase as is shown in \cite{usama}.
The MI pattern associated with the attractive interaction plays a key role in the formation of bright solitons in a pure BEC \cite {Sal, Car}. 
By investigating the time evolution of soliton trains, we find that the number of bright solitons is increased with increasing the impurity-boson interactions. 

This paper is organized as follows. In section \ref{flism}, we briefly review the main features of our theoretical approach. 
Section \ref{sol} is dedicated to analyze the behavior of a single soliton in the impurity component of a BEC-impurity mixture where  we solve analytically and numerically the 
generalized self-focussing nonlinear Schr\"o\-dinger equation.
Section \ref {ST} presents the generation strategy of bright soliton trains in a trapped BEC-impurity mixture.
We show in particular how the solitary wave formation occurs in the impurity through the MI.
Section \ref{concl} is devoted to conclusion.

\section{Formalism} \label{flism}

We consider a mobile impurity of mass $m_I$ immersed in a BEC of atoms of mass $m_B$ at finite temperature. 
The impurity-boson interaction and boson-boson interactions have been approximated by the contact potentials $g_B \delta ({\bf r}-{\bf r'})$ and $g_{IB} \delta ({\bf r}-{\bf r'})$, respectively. 
We neglect the mutual interactions of impurity atoms since we  assume that their number and local density remains sufficiently small \cite {Tim1, Jack1} 
and hence there is no impurity fluctuation. 
The TDHFB equations which govern the dynamics of the condensate, the thermal cloud, the anomalous density and the impurity read \cite {boudj2014, boudj12014}

\begin{subequations}\label{E:td}
\begin{align}
&i\hbar \dot{\Phi}_B = \left [h_B^{sp}+g_B \left((\beta-2) n_B+2n+\gamma n_I\right) \right]\Phi_B, \label{E:td1} \\ 
&i\hbar \dot{\tilde{m}} = 4\left[ h_B^{sp}+g_B \left(2G \tilde {m}+2n+\gamma n_I\right)\right]\tilde{m}, \label{E:td2}\\
&i\hbar \dot{\Phi}_I  = \left[ h_I^{sp} +g_{IB} (n_B+\tilde{n})\right]\Phi_I.  \label{E:td3} 
\end{align}
\end{subequations}
In the set (\ref {E:td}), $h_B^{sp}=-(\displaystyle\hbar^2/\displaystyle 2m_B) \Delta + V_B$ and $h_I^{sp}=-(\displaystyle\hbar^2/\displaystyle 2m_I)\Delta + V_I$ 
are, respectively the single particle Hamiltonian for the condensate and the impurity, with $V_B$ and $V_I$ being, respectively the condensate and the impurity trap potentials. 
$\Phi_B {(\bf r)}=\langle \hat\psi_B (\bf r)\rangle$  is the condensate wave function, $n_B{(\bf r)}={|\Phi_B(\bf r)|}^2$ is the condensed density, $
\Phi_I{(\bf r)}=\langle \hat\psi_I (\bf r)\rangle$ is the impurity wave function, $n_I {(\bf r)}={|\Phi_I (\bf r)|}^2$ is the density of impurity atoms, 
the noncondensed density $\tilde{n}(\bf r)$ and the anomalous density $\tilde{m}(\bf r)$ are identified respectively with 
${\langle \hat\psi^{+} {(\bf r)} \hat\psi (\bf r)\rangle }-\Phi_B ^{*} {(\bf r)} \Phi_B {(\bf r)}$ 
and ${\langle \hat\psi{ (\bf r)} \hat\psi (\bf r)\rangle}-\Phi_B {(\bf r)}\Phi_B {(\bf r)}$, where $\hat\psi^{+}$ and $\hat\psi$ 
are the boson destruction and creation field operators, respectively. The total density in BEC is defined by $n=n_B+\tilde {n}$. 
The dimensionless parameters $\beta = U /g_B$  with $U=g_B(1+\tilde {m}/\Phi_B ^2)$ being the renormalized coupling constant \cite {boudj2014, boudj12014},
$G = \beta / 4(\beta-1 )$ and $\gamma=g_{IB}/g_B$ is the relative coupling strength.
For $\beta = 1$, i.e., $ \tilde{m} /\Phi_B^2 = 0$, Eq.(\ref {E:td1}) reduces to the HFB-Popov
equation which is safe from all ultraviolet and infrared divergences and thus provides a gapless spectrum. 
For $0<\beta<1$, $G$ is negative and hence, $\tilde{m} $ has a negative sign. 
For $\beta >1$, $G$ is positive, and thus, $\tilde{m} $ becomes a positive quantity.
\\
Neglecting the mean-field interaction energy between bosons and impurity components i.e. $g_{IB} =0$ and setting $\tilde{n}=\tilde{m}=0$,
one recovers the well known Gross-Pitaevskii equation describing a degenerate Bose gas at zero temperature
and the Schr\"o\-dinger equation describing a noninteracting impurity system. 
For further computational details, see Refs.\cite {boudj2010, boudj2011, boudj2012, boudj2013, boudj2014, boudj12014}.

In our formalism the noncondensed and the anomalous densities are not independent. 
By deriving an explicit relationship between them, it is possible to eliminate $\tilde{n}$ via \cite{ boudj2014, boudj12014}:
\begin{equation}  \label{Inv}
I =(2\tilde{n} +1)^2-4|\tilde{m} |^2.
\end{equation}
One can easily check by direct substitution that once Eq.(\ref{Inv}) holds initially, it remains true during the dynamical evolution. 
The simplified set of equations are then the coupled equations (\ref{E:td}) with $\tilde{n}$ is replaced by the expression (\ref{Inv}).
In the uniform case, by working in the momentum space, $I_k=\coth^2 ( \varepsilon_k/T)$ \cite {boudj2012},  where $\varepsilon_k$ is the excitation energy of BEC.
At zero temperature $I=1$ \cite{boudj12014}, and hence, Eq.(\ref{Inv}) reduces to $\tilde{n} (\tilde{n} +1)=|\tilde{m} |^2$.
Therefore, the expression of $I$ clearly shows that $\tilde{m}$ is larger than $\tilde{n}$ at low temperature, so the omission of the anomalous density in this situation 
is principally unjustified approximation and wrong from the mathematical point of view.
Importantly, the expression of $I$ not only renders the set (\ref{E:td}) close but also enables us to reduce the number of equation making the numerical simulation easier.

Moreover, what is important in the TDHFB approach for Bose systems is that there have been
no assumptions on weak interactions. Therefore, the theory is valid even for strong interactions \cite{boudj2010,boudj2011}.
In addition, the TDHFB equations (\ref{E:td}) satisfy the total number of particles and the energy
conservation law, and they provide a gapless spectrum \cite{boudj12014}.

The stationary TDHFB equations (\ref {E:td}) can be easily obtained within the transformations: $\Phi_B(x,t)=\Phi_B (x)\exp (-i\mu_B t/\hbar)$ and
$\tilde{m} (x,t)=\tilde{m} (x)\exp (-i\mu_{\tilde{m}} t/\hbar)$, where $\mu_B$ and $\mu_{\tilde{m}}$ are, respectively, 
chemical potentials of the condensate and the anomalous density. 
If the condensed and the anomalous densities change smoothly we can apply the Thomas-Fermi
approximation, i.e. neglect the kinetic terms in Eqs. (\ref {E:td1}) and (\ref {E:td2}).
In this regime we immediately find that

\begin{equation}  \label{eq10}
n_B=\frac{1}{(\beta-2)} \left[\frac{\mu_B- V_B }{g_B}-2n-\gamma n_I\right],
\end{equation}
\begin{equation}  \label{eq11}
\tilde {m}=\frac{1}{2G} \left[\frac{\mu_{\tilde {m}}-V_B }{g_B}-2n-\gamma n_I\right].
\end{equation}
The density profiles (\ref {eq10}) and (\ref {eq11}) have the form of an inverted parabola. 
For $\gamma =0$, $n_B$ an $\tilde {m}$ reduce to their usual expressions. It is remarkable that for $\beta=1$, the anomalous density vanishes.\\

Assuming now that the impurity-boson coupling is sufficiently weak. Substituting Eqs.(\ref{eq10}) and (\ref{eq11}) in Eq.(\ref{E:td3}),
we then obtain the generalized self-focussing nonlinear Schr\"o\-dinger equation (NLSE)
\begin{equation}  \label{impur}
i\hbar \dot{\Phi}_I = \left( -{\frac{\hbar^2}{ 2m_I}}\Delta + V_I -\lambda V_B-\nu_I-g_B \gamma\lambda n_I\right)\Phi_I, 
\end{equation}
where $\lambda= \gamma [1/(\beta-2)+1/2G]$ and $\nu_I= 2ng_B\lambda-\gamma [\mu_B/(\beta-2)-\mu_{\tilde{m}}/2G] $.\\
The effective interaction potential $\lambda$ is positive only when the system is strongly correlated i.e for $\beta>2$ or $\tilde {m} > n_B$.
Whereas it is negative for $1<\beta<2$. Importantly, $\lambda$ has a minimum at $\beta=\beta_c=6.83$.\\
Equation (\ref{impur}) has never been obtained before in the literature. It allows us to study, in useful manner, the dynamical properties of the impurity
and the fluctuation effects of the impurity environment at the same time. For $\beta=1$, Eq.(\ref{impur}) reduces to the standard NLSE.


Let us consider now an impurity immersed in an elongated (along the $x$-direction) BEC and confined in a highly anisotropic trap 
(such that the longitudinal and transverse trapping frequencies are $\omega_{Bx}/\omega_{B\perp}\ll1$). 
In such a case, the system can be considered as quasi-1D and, hence, the coupling constants of equations (\ref {E:td}) effectively
take their 1D form, namely $g_B =2\hbar\omega_{B\perp} a_B$ and $g_{IB} =2\hbar\omega_{B\perp} a_{IB}$, where  $a_B$ and $a_{IB}$ are the scattering lengths describing
the low energy boson-boson  and impurity-boson scattering processes.\\
To well investigate the behavior of solitons in the impurity component, it is convenient to reformulate Eq.(\ref {impur}) in terms of dimensionless quantities using the following parameters:
$x=x/l_I$ where $l_I=\sqrt{\hbar/m_I\omega_I}$ is the impurity oscillator length, $\alpha=m_B/m_I$ is the ratio mass, $\Omega_{\perp}=\omega_{B\perp}/\omega_{I\perp}$, 
with $\omega_{I\perp} $ being the transverse impurity confinement frequency, $\tau=t\omega_{I\perp}$ and ${\Phi}_I=\Phi_I l_I^{1/2}$. \\
Equation (\ref {impur}) turns out to be given as 
\begin{equation}  \label{impurT}
i\frac{d{\Phi}_I}{d\tau} = \left( -{\frac{1}{2}}\Delta_x +\frac{1}{2} \varrho^2 x^2-\bar\nu_I-\bar g n_I\right) {\Phi}_I, 
\end{equation}
where $\varrho^2=1-\lambda\alpha\Omega_{\perp}^2$, $\bar\nu_I=\nu_I/\hbar\omega_{I\perp}$ and $\bar g=2\lambda \Omega_{\perp} (a_{IB}/l_I)$.\\
The energy functional corresponding to the NLSE (\ref{impurT}) reads
\begin{equation}\label{enr}
E=\int_{-\infty}^\infty \left[\frac{1}{2} |\nabla\Phi_I|^2+\left(\frac{1}{2}\varrho^2x^2-\bar\nu\right) |\Phi_I|^2-\frac{\bar g}{2}|\Phi_I|^4 \right]dx.
\end{equation}
Remarkably, for strong pair correlations where $\beta>2$,  $\bar g >0$, the impurity wave packet has an attractive interaction between atoms, 
so that the system supports automatically bright solitons. For $1<\beta<2$, the impurity behaves as a dark soliton.\\

\section{Single solitons} \label{sol}
In this section,  we restrict ourselves to study the properties of single bright solitons. 

In the static case and for $\varrho=0$, Eq.(\ref{impurT})  is integrable and admits the solution 
\begin{equation}\label{StatSol}
{\Phi}_I (x)=\frac{1}{\sqrt {2\Lambda}} \,\text{sech} \left(\frac{x}{\Lambda}\right),
\end{equation}
where $\Lambda=2/\bar g$ is the extended localization length.
The energy of the bright soliton is calculated through Eq.(\ref{enr}) as $E=-\bar g ^2/24$.

To analyze the time evolution of the single bright soliton in detail, we fully solve Eq.(\ref{impurT})  numerically employing the finite-difference splitting scheme.
\begin{figure}
  \centering{
  \includegraphics[scale=.8, angle=0]{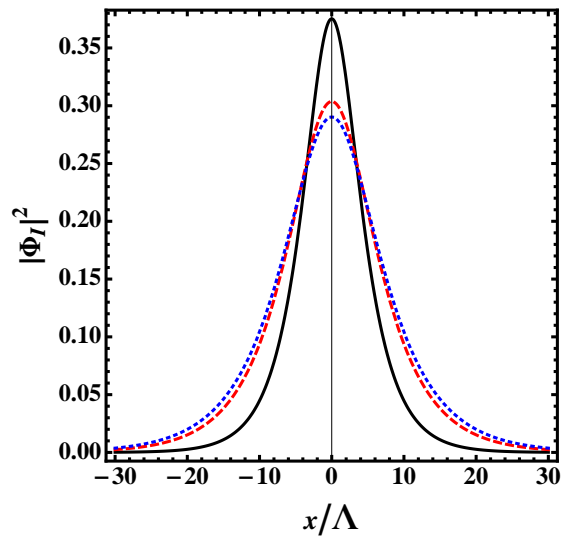}}
  \caption{(Color online) Density profile of bright soliton in the impurity component.
The parameters are set to: $N_I=1$ of ${}^{85}$Rb impurity atom, $N$=$10^5$ of ${}^{23}$Na bosonic atoms, $a_B$=3.4nm, $a_{IB}$=16.7nm, 
and the transverse trapping frequencies are $\omega_{B \perp}=\omega_{I \perp}=2\pi \times500$ Hz.
Blue dotted line: $\beta=\beta_c$, red dashed: line $\beta$=3.5 and solid line:$\beta$=2.5.}
  \label{sol1}
\end{figure}
Figure. (\ref{sol1}) depicts that the soliton density decreases continuously until it reaches its minimal value
for $\beta \approx \beta_c$ or equivalently $\tilde{m}=(\beta_c-1) n_B \approx 5.83n_B $. 
This latter condition which determines the soliton stability threshold, clearly indicates that the condensate is dominated by its anomalous fluctuations.
As a result, a stable and robust impurity bright soliton locates at the local minimum of the effective potential created by the impurity-host interaction. 
The decay  arises from the fact that at nonzero temperatures the condensate coexists with both a noncondensed cloud and anomalous density composed of thermally excited
quasiparticles. 
Therefore, interactions between condensed and noncondensed atoms on the one hand and interactions of the impurity soliton with atoms 
of the surrounding bath on the other hand lead to dissipation, and hence the soliton loses energy.
Indeed, the effect of the interaction parameter $\beta$ has considerable consequences, not only on the structure
of the soliton, but also on the dynamics of the system, as we will see later on.

\section {Soliton Train}\label {ST}
We investigate in this section the formation and the dynamics of bright soliton trains in the impurity component of harmonically trapped BEC-impurity mixture.

Let us start first by illustrating the effect of the MI on the nonlinear evolution of the impurity.
For a given stationary solution $\Phi_{I0}$ the NLSE (\ref{impurT}) with eigenvalue $\mu_I$,
the small-amplitude excitations of the system are defined through the random phase approximation (RPA) as
$\Phi_I=\Phi_{I0}+\delta\Phi_I$ with $\delta\Phi_I=[u_q (x) e^{-i\omega_q \tau }+v_q (x) e^{i\omega_q \tau} ] e^{-i\mu_I \tau}$. 
We then obtain
\begin{equation} \label{BdG}
\begin{pmatrix} 
{\cal L} & {\cal M} \\
{\cal M} & {\cal L}
\end{pmatrix}
\begin{pmatrix} 
u_q(x) \\ v_q(x)
\end{pmatrix}=\omega_q 
\begin{pmatrix} 
u_q(x) \\- v_q(x)
\end{pmatrix},
\end{equation} 
where ${\cal L}=-\Delta/2 +\varrho^2 x^2/2 -\bar\mu - 2\bar g n_I$ with $\bar\mu=\bar\nu+\mu_I$ being the chemical potential of impurity soliton,
 ${\cal M}=-\bar g\Phi_I^2$ and $ u_q (x), v_q (x)$ 
are the quasi-particle amplitudes.\\
In the case of modulational instability analysis of homogeneous quasi-1D case, the coupled system (\ref{BdG}) is solved with  being proportional to
a traveling plane wave solution of wave number $k$ and complex dispersion $\omega_q$. 
Thus resulting in the following dispersion relation
\begin{equation} \label{dis}
\omega_q= \sqrt{\frac{q^2}{2} \left(-\frac{q^2}{2}+2\bar g n_I\right)}.
\end{equation} 
For $\bar g<0$, the value of $\omega_q$ is always positive, whereas for $\bar g>0$, the frequency $\omega_q$ has an imaginary part 
and the spatially modulated perturbations grow exponentially with time, as is evident from the form of $\delta\Phi_I$.
Therefore, the MI occurs for repulsive coupling constant $\bar g$ unlike to the case of a pure BEC. 
This can be attributed to the effects of the anomalous correlations which are strong enough to change the sign of the interactions.
The fastest growth occurs for the wave number $q_{max}$ that gives a maximum of $\omega_q$. The extremum condition $\partial \omega_q/\partial q=0$
gives $q_{max}=\sqrt{2\bar g n_I}$ and the maximum rate of growth is $q_0=q_{max}/\sqrt{2}$.
In the context of the nonlinear optics, MI occurs also for any sign of nonlinearity and GVD when the electromagnetic field is polarized, 
the light propagation in isotropic Kerr media is described by two incoherently coupled NLSE which referred to as cross-phase-modulation, 
and thus leads to MI for any sign of the interaction strength \cite{Agr}. 



\begin{figure}
  \centering{
  \includegraphics[width=4.2cm,height=4cm]{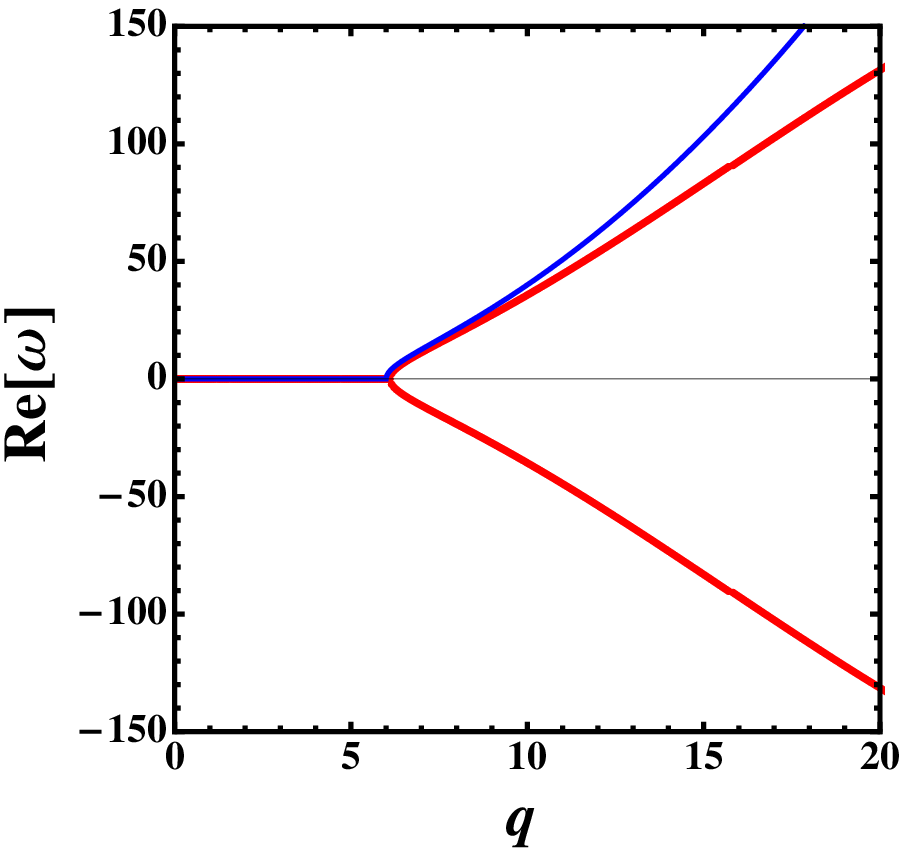}
  \includegraphics[width=4.2cm,height=4cm]{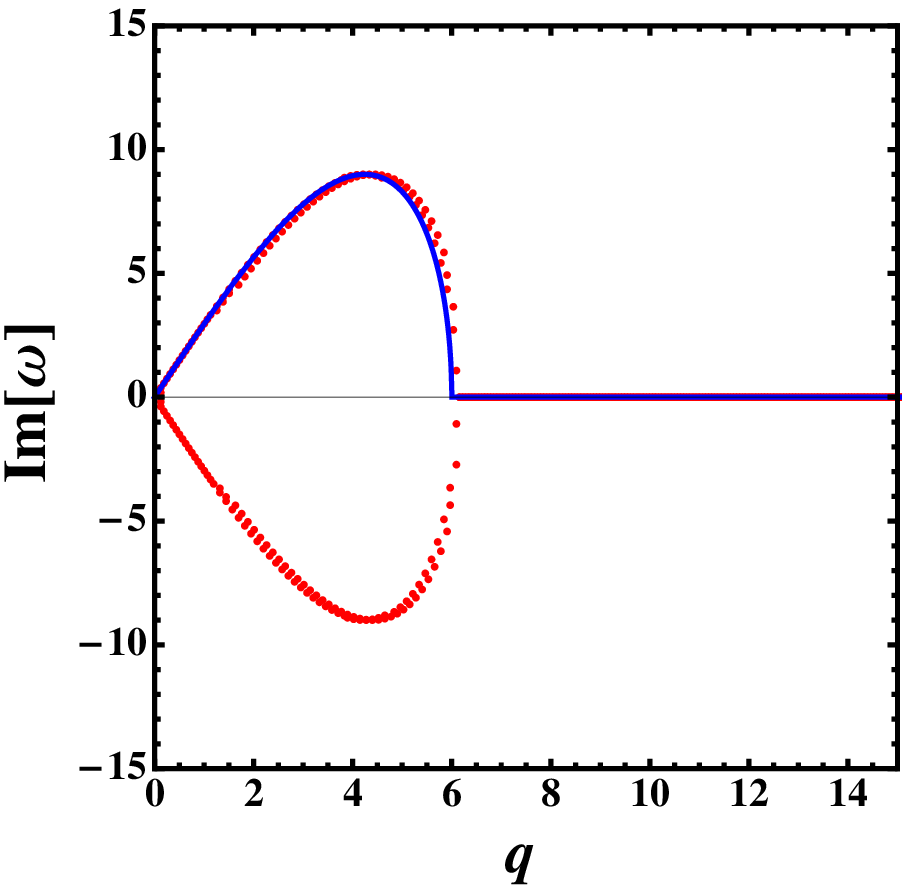}}
  \caption{ The real and imaginary parts of the unstable eigenvalue pertaining to the bright soliton versus the wave number of the corresponding eigenmode 
calculated by diagonalizing Eq. (\ref{BdG}) (red circles). The blue curves correspond to the uniform case as given by Eq. (\ref{dis}). Parameters are the same as in Fig. \ref{sol1}.}
  \label{MI1}
\end{figure}

In the trapped case, the dispersion relation can be obtained by taking the Fourier transform of the wave number associated with each eigenmode calculated by
diagonalizing the set (\ref{BdG}). The dispersion spectrum $\omega_q$ is displayed in Fig.\ref{MI1}.
Plotting the analogous curves for the uniform case, where Eq. (\ref{dis}) provides analytical expression, 
we observe an excellent agreement  for the real part of the spectra while a small deviation
appears in the imaginary part of the spectra at larger $q$. 

An important consequence of the MI is the formation of trains of bright solitons in the impurity component of a trapped BEC-impurity mixture.
To investigate the generation of these trains, we first use the above ground-state wave function  (\ref{StatSol}) as an initial condition for the time-evolution of Eq. (\ref{impurT}).
The numerical code implements again in cylindrical symmetry a finite-difference splitting method.
Assuming that the system is in equilibrium i.e. $\beta =\beta_c$ (to avoid dissipation effects) and increasing the strength of the coupling
between bosons and impurity $a_{IB}$.

\begin{figure}
  \centering
  \includegraphics[width=8 cm,height=10cm]{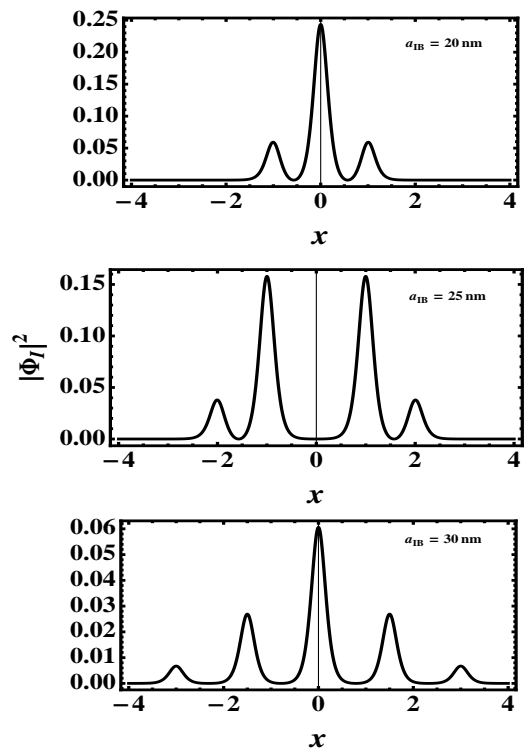}
  \caption{ Axial density profile of the BEC made for $\beta=\beta_c$. Parameters as in Fig.\ref{sol1}.}
  \label{Train}
\end{figure}

Figure. \ref{Train} shows that at $\tau=5.4$, the impurity cloud of ${}^{85}$Rb-${}^{23}$Na mixture breaks into several peaks which remain 
spatially localized during the time evolution and thus forming soliton trains. The number of peaks grows with the scattering length $a_{IB}$. 
The formation of these trains can be explained as due to the strong anomalous correlations which grow themselves with increasing interactions \cite{boudj2011}.
Note that the pairing correlations lead also to split the bright soliton into two partially coherent solitonic structures of opposite momenta in a pure BEC \cite{Bulj}.
Furthermore, the bright soliton trains seen in Fig.\ref{Train} generate by virtue of the MI of the time-dependent wave function of the impurity, 
driven by imaginary Bogoliubov excitations as we have mentioned above.

\section{Conclusion} \label{concl}

By using the TDHFB theory we have shown that bright solitons can be generated in the impurity component of BEC-impurity mixture at finite
temperature as a result of strong anomalous correlations. 
We have found that these solitons decay due to the impurity-medium and medium-medium interactions.
We have shown that after a sudden increase of the impurity-boson strength a train of bright solitons is produced 
in the impurity component and this can be interpreted in terms of the MI of the time-dependent impurity wave function. 
In our case the MI occurs even for repulsive interactions which is a salient feature in  BEC-impurity mixtures.
Most recently, the formation of bright solitons with repulsive nonlinearity 
has been observed from the surface spin-wave propagating in yttrium iron garnet thin film strips \cite {Zih}. 
Moreover, we have pointed out that the anomalous correlations, which become more and more stronger with increasing $a_{IB}$, 
is also an important ingredient for the generation of bright soliton trains.
The number of solitons in the train is proportional to $a_{IB}$.		


\section{Acknowledgements}

We are grateful to Usama Al Khawaja and Yuri Kivshar for comments on the manuscript.
This work was supported by the Algerian government under Research Grant No. CNEPRU-•	B00L02UN020120130004.

\end{document}